# From Protest to Power Plant: Interpreting the Role of Escalatory Hacktivism in Cyber Conflict

Richard Derbyshire[a]*, Diana Selck-Paulsson[a], Charl van der Walt[a], and Joe Burton[b]

[a] *Security Research Center, Orange Cyberdefense*

[b] *Department of Politics, Philosophy and Religion, Lancaster University, United Kingdom*

Ric Derbyshire is a Principal Security Researcher at Orange Cyberdefense and an Honorary Researcher at Imperial College London. He holds a PhD in computer science from Lancaster University, where he researched adversary-centric quantitative risk assessment. His work takes a pragmatic approach to both offensive and defensive aspects of cyber security, with a particular focus on operational technology and critical national infrastructure.

Diana Selck-Paulsson is a Senior Security Researcher for Orange Cyberdefense, where among other projects she heads the company's efforts to track, analyse and report on the current Cyber Extortion (Cy-X) / Ransomware threat landscape, as well as hacktivism activities since 2022, when the lines between different forms of cyber aggressions began to blur. Diana has a master's in international criminology. Her primary interests and focus are on data analysis, researching cybercrime trends, victimology and the human element in cyber security.

Charl van der Walt is a South African cyber security expert with over 25 years' experience. He co-founded SensePost, a pioneering penetration testing firm, and led it until its acquisition by SecureData. Today, he is Global Head of Security Research at Orange Cyberdefense, where he leads a team analysing and communicating emerging security threats to inform strategy and advisories. An internationally recognized speaker at events such as RSAC and Black Hat, Charl is valued for translating complex research into practical security insight.

Dr Joe Burton is Professor of Security and Protection Science at Lancaster University. He is the author of NATO's Durability in a Post-Cold War World (SUNY Press, 2018), editor of Emerging Technologies and International Security: Machines the State and War (Routledge, 2020), and his work on AI and Cyber Security has been published in a range of leading scientific journals, including International Affairs, Journal of Global Security Studies, Technology in Society, Asian Security, Defence Studies, the Cyber Defence Review, the RUSI Journal and Political Science.

# From Protest to Power Plant: Interpreting the Role of Escalatory Hacktivism in Cyber Conflict

Since 2022, hacktivist groups have escalated their tactics, expanding from distributed denial-of-service attacks and document leaks to include targeting operational technology (OT). By 2024, attacks on the OT of critical national infrastructure (CNI) had been linked to partisan hacktivist efforts in ongoing geopolitical conflicts, demonstrating a shift from protest to something more resembling cyber warfare. This escalation raises critical questions about the classification of these groups and the appropriate state response to their growing role in destabilizing international security.

This paper examines the strategic motivations behind escalatory hacktivism, highlighting how states may tolerate, encourage, or leverage hacktivist groups as proxies in conflicts that blur the lines between activism, cybercrime, and state-sponsored operations. We introduce a novel method for interpreting hacktivists based on the impact of their actions, alignment to state ideology, and host state involvement, offering a structured approach to understanding the phenomenon. Finally, we assess policy and security implications, particularly for host and victim states, and propose strategies to address this evolving threat. By doing so, this paper contributes to international discussions on cyber security policy, governance, and the increasing intersection between non-state cyber actors and state interests.

## Introduction

Since 2022, cyber-attacks by hacktivist groups have surged, closely linked to the onset of the war in Ukraine and shifting global conflict dynamics (Zafra, et al., 2024). Historically, hacktivism has centred on ideals like free access to information, privacy rights, and ethical technology use, or has been reactionary, targeting establishments. However, in recent years, it has increasingly adopted a geopolitical dimension, with some groups aligning with state-backed narratives and taking sides in international conflicts (International Committee of the Red Cross, 2024). This marks a departure from hacktivism's idealistic, independent origins, reflecting its growing entanglement with geopolitical agendas. As a result, hacktivism has become increasingly distinct from its original form.

Accompanying this ideological transformation is an escalation in the scope and sophistication of tactics employed. Hacktivist activities now extend beyond traditional cyber disruptions, such as distributed denial of service (DDoS) attacks, to targeting operational technology (OT), bridging cyber and physical domains. Early OT attacks were often isolated to manufacturing facilities, but recent developments include targeting critical national infrastructure (CNI). These attacks impact essential services like energy, water, and transportation, directly affecting civilians in victim states. The deliberate focus on CNI signals increased strategic intent within hacktivist operations, amplifying their real-world consequences.

The evolution of hacktivism marks a shift in the cyber landscape, transforming it from an idealistic phenomenon into a complex extension of global power struggles. As hacktivist groups align more closely with state narratives, the line between independent activism and state-sponsored operations has blurred, raising critical questions about how to address modern hacktivism. This paper contributes to understanding this shift by introducing the Cyber Impact-Alignment-Responsibility Spectrum (CIARS), a novel method to categorize hacktivist activities based on the impact of their attacks, their alignment with their host state's ideology, and the involvement of their host state. In turn, this facilitates a better understanding of the implications of escalatory hacktivism and how to address it.

The remainder of this paper is structured as follows: Section 2 reviews the historical evolution of hacktivism, focusing on the rise of pro-establishment hacktivism and its challenges. Section 3 examines existing interpretations of hacktivist groups and introduces a novel multi-dimensional spectrum with supporting use cases. Section 4 explores the implications of escalatory hacktivism for victim states, including disruptions to critical national infrastructure (CNI) and destabilisation risks, and for host states, which face challenges such as reputational damage, diplomatic fallout, and questions of state responsibility. Section 5 considers policy responses and strategies to address these challenges. Finally, Section 6 concludes with key findings.

## Background

This paper focuses on the interpretation, implications, and responses to the contemporary hacktivist threat, a focus shaped by the escalation and intensification of hacktivist tactics in recent years. To understand this shift, it is helpful to briefly examine the evolution of hacktivism across its previous eras. This historical context forms a basis for comparing past motivations, tactics, and targets with those of today, shedding light on the challenges posed by modern hacktivism.

Hacktivism's history can be categorized into three distinct eras: the Digital Utopia Era, the Anti-Establishment Era, and the Establishment Era (Selck-Paulsson & Gibney, 2024). These eras are defined by the evolving ethos of hacktivist groups and their corresponding tactics. While a new era does not erase the characteristics from its predecessors, the influence of prior ethos and tactics typically diminishes as those of the contemporary era take precedence.

### *The Digital Utopia Era*

The first era of hacktivism emerged in the mid-1980s, defined by principles such as free access to information, privacy rights, and exposing security vulnerabilities in commercial software to protect end users. Cult of the Dead Cow (cDc) is widely recognized as the pioneer of hacktivism and the ethos of the Digital Utopia Era (Menn, 2019). However, similar movements soon appeared in Europe, particularly in Germany, with groups like Bayrische HackerPost (BHP) and the Chaos Computer Club (CCC) leveraging the Internet to disrupt systems and educate the public on these shared ideals (Chaos Computer Club, n.d.).

In the late 1990s, the Legion of the Underground (LoU) emerged as a more politically focused hacktivist group, targeting oppressive regimes and institutions, and even declaring war on Iraq and the People's Republic of China. While such aggressive, politically charged actions might be expected from hacktivist groups today, they were controversial at the time. This led other prominent groups, including cDc and CCC, to issue a joint statement condemning LoU (2600; Chaos Computer Club; Cult of the Dead Cow; !HISPAHACK; L0pht Heavy Industries; Phrack; Pulhas, 1997).

### *The Anti-Establishment Era*

With a gradual shift from the ethos of the Digital Utopia Era, the Anti-Establishment Era emerged around the mid-2000s. Prominent groups such as Anonymous, Wikileaks, and Lulzsec though differing in principles, shared a cynicism aimed at disrupting governments and corporations (Goode, 2015). Most hacktivists of this era abandoned aspirations for a digital utopia, becoming primarily reactionary by targeting social injustices, censorship, and systemic oppression. Some actions directly retaliated against geopolitical conflicts, while others were carried out simply 'for the lulz' (Karagiannopoulos, 2021).

### *The Establishment Era*

The transition from the Anti-Establishment Era to the Establishment Era marked an about-turn in the ethos of hacktivist groups. In this new era, many hacktivists began aligning with or conducting campaigns on behalf of governments, religious institutions, and nation-states. Anti-establishment activities, such as anti-war protests, gave way to pro-establishment operations supporting either side of geopolitical conflicts. The relationship between states and hacktivist groups in this era can vary significantly. It may take the form of co-option, where states directly influence or coordinate with these groups, or passive toleration, where states allow their activities to proceed unchallenged to serve strategic objectives (Maurer, 2018). This shift was particularly evident during Russia's 2014 military intervention in Ukraine, when hacktivist groups on both sides claimed responsibility for cyber-attacks (Maurer, 2015).

After 2014, hacktivism experienced a period of relative quiet, as its activities simmered and further entrenched the ethos of the Establishment Era (IBM, 2019). During this time, groups like the Syrian Electronic Army (SEA) carried out various cyber-attacks explicitly aligned with the interests of the Assad regime, targeting opposition groups and critical media outlets (Caldwell, 2015) (Bertram, 2017). By 2019, however, hacktivism resurged, with Anonymous conducting state-aligned 'operations' in support of pro-democracy movements in Hong Kong and later extending efforts to Ukraine amidst geopolitical tensions (Betlej, 2023).

*The Escalation and Intensification of Hacktivism Tactics*

In 2022, hacktivist groups began escalating their tactics, possibly driven by the emergence of various geopolitical conflicts. Unlike earlier eras, when hacktivists positioned themselves in opposition to state power, they now openly aligned with national objectives. That openness legitimized their actions and the narratives they projected publicly, recasting them from protestors to combatants in national or ideological conflicts. Evidence for this shift lay in their consistent self-attribution of attacks to their allegiance with one side or another. This reflected an escalation in a non-technical sense, less about tools and more about the political legitimacy they claimed.

A key technical development in escalatory hacktivism was the targeting of operational technology (OT), reflecting a similar trend among cyber criminals (Derbyshire, Green, Van der Walt, & Hutchison, 2024). While many claims by hacktivist groups remain unverified or exaggerated, there have been instances of successful cyber-physical attacks (Zafra, Lunden, & Brubaker, 2023). Successfully targeting OT is particularly significant, not only due to its cyber-physical impact but also because developing the requisite capabilities within an emergent and niche cyber security context is costly for any adversary group lacking external support, such as that provided by a state (Derbyshire, 2022).

One notable example of escalatory hacktivism was Predatory Sparrow, a purportedly pro-Israel group, which in June 2022 conducted cyber-physical attacks on the OT of three Iranian steel manufacturers, causing kinetic impacts. The group cited unspecified aggression by the Islamic Republic as justification. They emphasized that no casualties occurred, claiming they deliberately executed the attacks carefully to protect civilians and warned Iran's emergency services in advance (Kalinowski, 2022). Fortunately, the attacks targeted isolated manufacturing facilities, minimizing broader impacts on civilian populations.

In 2023, CyberAv3ngers, a hacktivist group reportedly aligned with Iran and linked to the Iranian Government's Islamic Revolutionary Guard Corps (IRGC), intensified hacktivist tactics by targeting OT within critical national infrastructure (CNI). Motivated by anti-Israel sentiment, the group attacked Unitronics Vision Series programmable logic controllers (PLCs) and human-machine interfaces (HMIs). The attacks caused moderate disruption, including HMI defacements with anti-Israel messaging, forcing one facility to operate manually and leaving another unable to supply water to 180 customers for two days (CISA, 2024).

In 2024, the pro-Ukrainian hacktivist group OneFist claimed on X (formerly Twitter) to have conducted a campaign against Russian power plants. The group shared images and videos of HMI mimics displaying live process data, implying they had manipulated those HMIs to disrupt power generation at targeted facilities (OneFist, 2024).

These attacks have demonstrated a trajectory of escalation and intensification, often aligning with state-backed narratives in conflicts. In certain cases, these operations appear strategically relevant to the states they support. Furthermore, while unverified at the time of writing, varying degrees of state involvement have been suggested (CISA, 2024). Over time, hacktivism has become increasingly distinct from its origins. To address this evolution, a structured approach is needed to systematically capture and analyse its nuances.

## Decomposing, delineating, and interpreting hacktivism

As outlined in Section 2, hacktivism has traversed a diverse range of motivations and impacts. Existing frameworks for understanding its implications or informing responses are outdated given the rapid escalation and intensification of tactics by contemporary hacktivist groups. A modern approach is therefore needed to address the complexities, nuances, and multi-dimensional nature of hacktivism and its escalation.

### *Existing Approaches to Interpreting Hacktivism*

The complexity and multiple dimensions of hacktivism are generally acknowledged in literature. However, there is often a disproportionate focus on individual dimensions in work attempting to categorize it. Many contributions categorize hacktivists by their primary ideology or motivation, such as the cause driving the hackers (Chng, Lu, Kumar, & Yau, 2022). Others focus on methods or strategies used and their legality or disruptiveness (Samuel, 2004). And yet others concentrate on organizational structure, considering forms from ad-hoc networks to more structured groups (Romagna M. , 2024) (Romagna & Leukfeldt, 2024).

Contemporary literature often focuses on the Establishment Era, emphasizing the interaction between states and hacktivists. This leads to reframing hacktivists as extensions of state power. For example, Svantesson (2023) uses the term 'cyber militia' to describe volunteers defending states with formal recognition or coordination, while Maurer (2018) categorizes hacktivists as 'cyber proxies' for their affiliated states. Similarly, Zettl-Schabath (2023) labels hacktivists as 'proxies' when their actions align with state objectives. This trend was reinforced by Pat McFadden, the UK's Chancellor of the Duchy of Lancaster, who, in a 2024 NATO Cyber Defence Conference speech, described Russia's cyber criminals and hacktivists as an 'unofficial army' (McFadden, 2024). Healey's (2011) spectrum of state responsibility captures this focus on state involvement effectively, ranging from state-prohibited to state-integrated, with intermediary categories such as coordinated, encouraged, and ignored.

Moving from state-involvement, Smith and Dean (2023) use the term 'digital resistance movement' to describe the IT Army of Ukraine, a hacktivist-like volunteer group organized by the Ukrainian Ministry of Digital Transformation in response to Russia's invasion. In contrast, Denning (2001) applies 'cyber terrorist' to politically motivated hacktivists causing violence against noncombatants or serious harm like loss of power or water, such as an attack on CNI.

Most literature provides or reinforces terms for hacktivism based on characteristics, motives, or eras, requiring justification for fitting groups into specific categories. Moreover, the recent focus on state involvement risks overemphasizing a single dimension while overlooking both the difficulties of determining it and the significance, nuance, and interaction of other dimensions

Three dimensions stand out as critical to interpreting hacktivist groups and their attacks: attack impact, state alignment, and state involvement. *Attack impact* implies the type and severity of a hacktivist attack. For instance, a DDoS of a small ecommerce website does not warrant the same response as disrupting CNI. *State alignment* examines whether a hacktivist group's actions align with a state's narrative, ideology, and objectives, regardless of the state's consent. *State involvement* considers whether a state can be implicated through direct support or tacit endorsement of hacktivist activities. To fully interpret hacktivist groups, understand their implications, and address their challenges, an approach is needed that considers the dynamic intersection of all three dimensions.

### ***The Cyber Impact-Alignment-Responsibility Spectrum***

Figure 1 depicts the Cyber Impact-Alignment-Responsibility Spectrum (CIARS). Rather than reverse engineer a hacktivist's impacts, alignment, and state involvement into a single definition, a term or range of terms can be ascertained with a combination of alignment and impact. Further knowledge of a hacktivist's relationship with their host state can be used, if available, to inform the user of whether the state should potentially be considered responsible for the impacts. To provide an understanding of CIARS' various axes and mechanisms, they are explained in the following subsections.

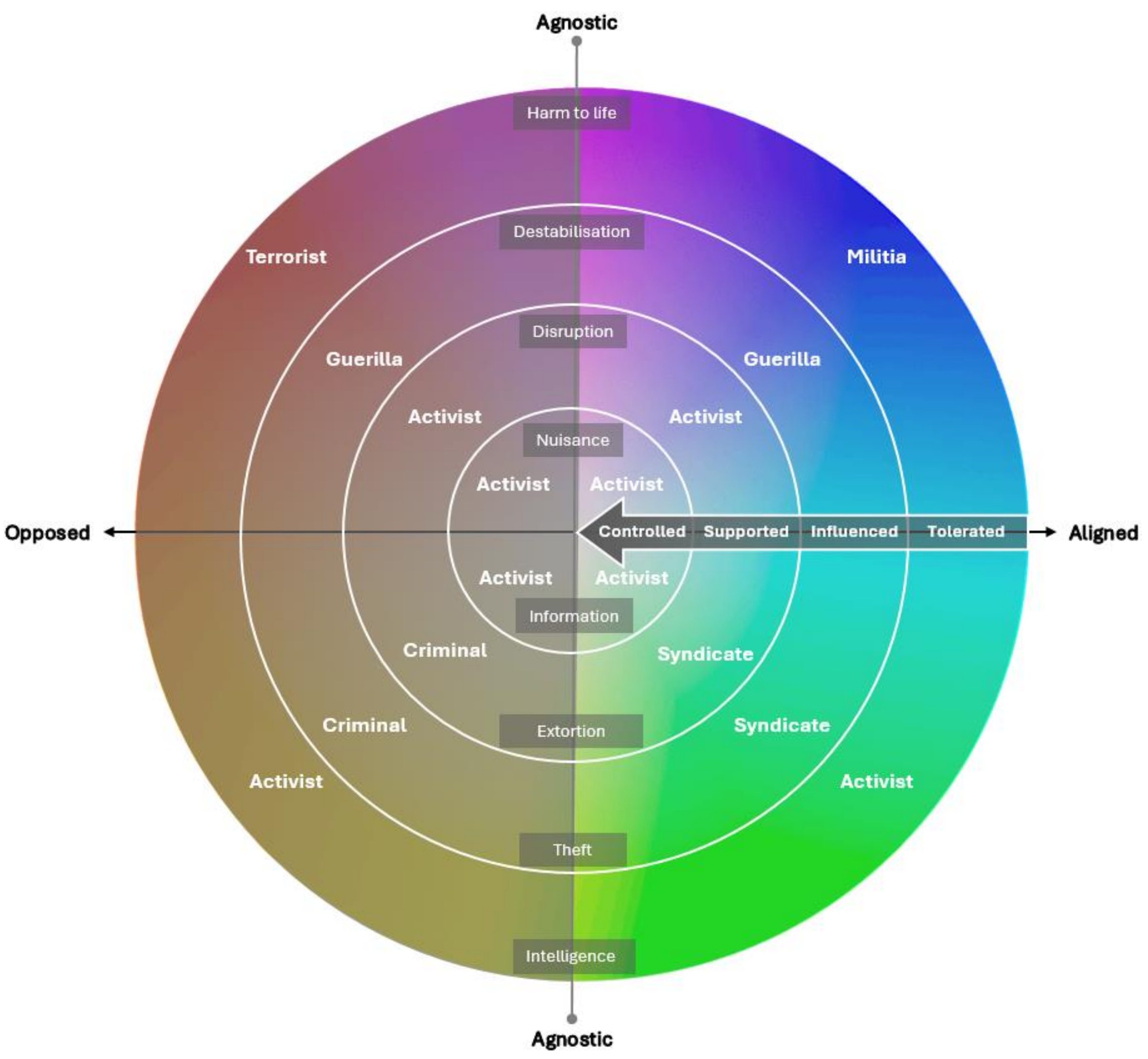

Figure 1. The Cyber Impact-Alignment-Responsibility Spectrum (CIARS)

*The Politico-Ideological Alignment Continuum*

The horizontal axis represents the degree of politico-ideological alignment between a hacktivist and its home state's objectives. It ranges from 'opposed', where the hacktivist resists the state's goals, to 'aligned', where the hacktivist's actions and ideology fully align with the state. The midpoint, 'agnostic', denotes hacktivists whose motivations are unrelated to the state's objectives. This continuum captures subtle variations in alignment levels.

*The Impact Continuum*

The vertical axis represents the range of impacts, arranged on a continuum delineated by the concentric circles within the spectrum. The centre line separates 'below the line' impacts, which are routine and non-escalatory, from 'above the line' impacts, which become more overt and escalatory. The further above the line an impact lies, the greater the escalation, while impacts far below the line are less so. Intuitive or widely understood terms are used where possible to guide users, but they should be used flexibly in extreme or anomalous cases that do not fit the existing descriptions.

The impacts above the centre line are defined as follows: *nuisance*, low-level effects causing annoyance or inconvenience, such as website defacements, low-grade DDoS attacks, or propaganda dissemination; *disruption*, moderate impacts interrupting critical services without widespread harm, such as financial outages, transport disruptions, or communication blackouts; *destabilization*, high-level impacts targeting systems critical to societal stability, causing widespread panic or economic harm, such as prolonged energy grid failures or significant healthcare disruptions; and *harm to life*, the most severe impacts, including loss of life or physical harm, such as attacks on safety-critical systems or industrial sabotage leading to casualties.

Below the centre line, impacts are categorized as follows: *information*, involving operations that manipulate or distort perceptions, such as disinformation, propaganda, and media manipulation; *extortion*, coercive activities using threats like exposing sensitive data, encrypting systems, or disrupting services to demand payment or compliance, including ransomware and sextortion; *theft*, unauthorized acquisition of assets for financial or strategic gain, such as credential theft, intellectual property theft, embezzlement, or fraud; and *intelligence*, covertly gathering sensitive information for economic, political, or military advantage, including espionage, reconnaissance, and surveillance.

*Impact-Alignment Hacktivist Classifications*

Hacktivists are classified by the impact of their attacks on the impact continuum, their state alignment along the politico-ideological continuum, and their distance from the spectrum's centre. These fluid classifications enable nuanced distinctions. Rather than introducing new terms, CIARS employs existing ones to connect emerging cyber behaviours with established geopolitical concepts, providing policymakers with familiar precedents for addressing challenges. The names are not prescriptive but suggest what term a hacktivist may fall under based on the impact and alignment observed. Superimposing these terms onto CIARS retains existing actor descriptions while incorporating contextual nuances through their relative positions. This approach offers a dynamic yet grounded understanding of hacktivists. By situating actors within these intersecting continua, CIARS makes it possible to compare hacktivists not only against each other but also against state and non-state actors in adjacent domains. In this way, the spectrum provides greater analytical precision than earlier categorical approaches.

*Incorporating State Responsibility*

CIARS focuses on its dual axes of politico-ideological alignment and impact. However, as Healey (2011) states, 'For national security policymakers, knowing "who is to blame?" can be more important than "who did it."' Therefore, incorporating the host state's responsibility for a hacktivist's actions, if known, is essential for shaping effective policy responses.

On the right side of CIARS, where hacktivists align with a state, terms are provided based on Healey's (2011) spectrum of national responsibility for cyber-attacks. At each ring, which implies a particular intersection of impact severity and politico-ideological alignment, a threshold of state involvement is provided. Should the state meet that, it is suggested they be considered responsible for the hacktivist's actions.

The *tolerated* category represents the lowest level of state involvement, where the state bears responsibility for failing to prevent or address the actions of *activist and militia actors*. These activities persist due to weak enforcement or deliberate oversight. The *influenced* category reflects a moderate level of involvement, where the state indirectly shapes or encourages the actions of *guerrilla and syndicate actors*, exerting influence without explicit direction. In the *supported* category, the state actively participates by providing resources or guidance, directly enabling the operations of *activist and syndicate actors* and facilitating their activities. Finally, the *controlled* category represents the highest level of involvement, where the state directly manages hacktivist actions through *execution*, *integration*, or *ordering*, fully assuming responsibility for their outcomes.

### ***CIARS Usage and Use Cases***

Plotting a hacktivist's actions on CIARS is a simple process of identifying approximate x/y coordinates based on each action's politico-ideological and impact alignment. These coordinates can then be mapped as they are or combined in a range, for a set of actions. It is important to view the impacts and classifications as suggestive rather than prescriptive, as over-interpretation may risk obscuring the fluidity and nuance the spectrum is designed to capture. As mentioned, the impacts should be considered flexible to account for anomalous or extreme cases, and the classifications are suggested terms that may or may not apply to a hacktivist in that area.

*Use Cases*

Figure 2 depicts four use cases of hacktivist groups mapped onto CIARS, with a key below it to denote their colour.

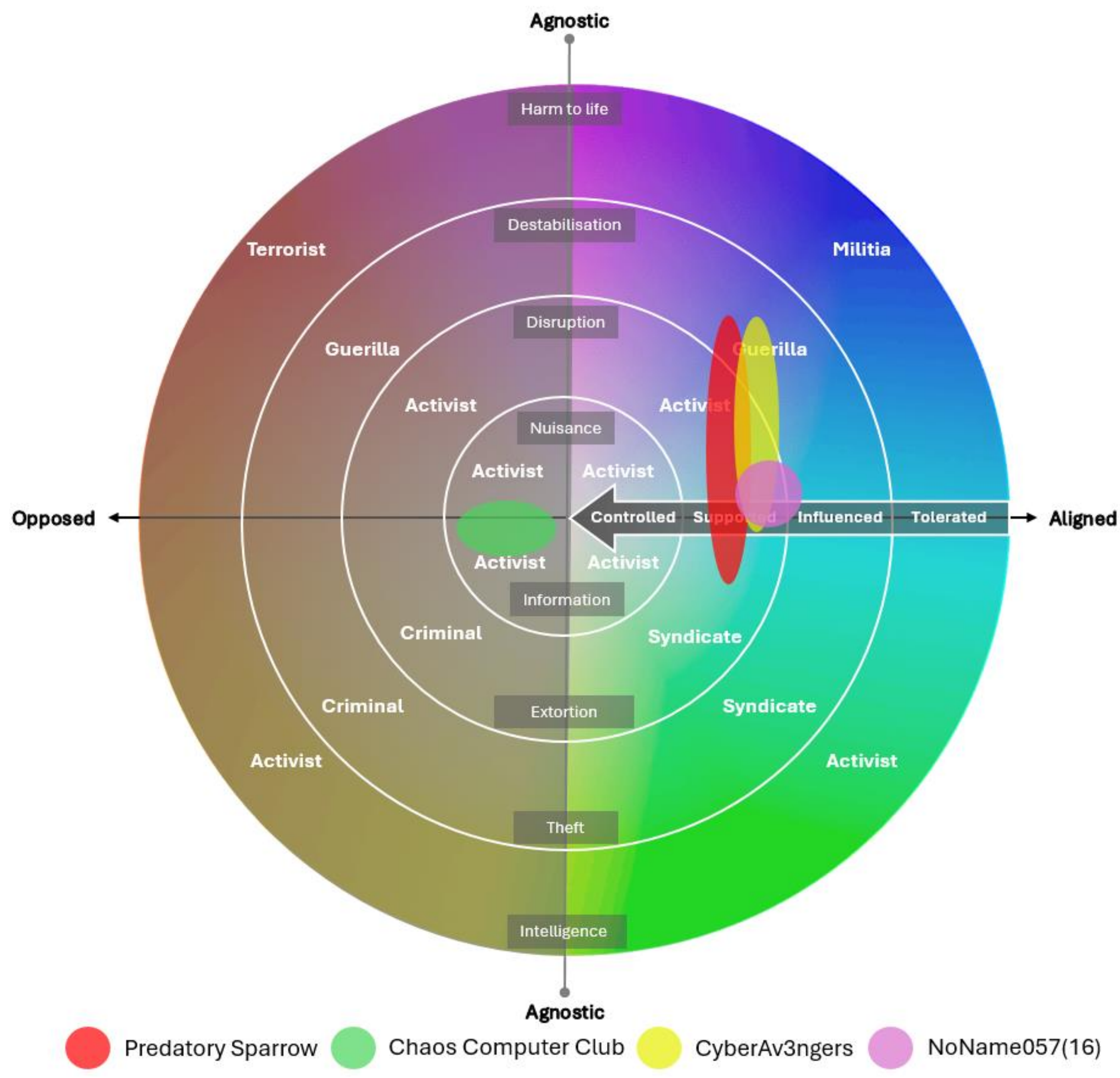

Figure 2. Example CIARS mappings with four use cases

As mentioned, *Predatory Sparrow* (PS) is a pro-Israel group, shifting their alignment right. Their activities range from information operations to attacks on Iranian steel manufacturers. While the latter does not align perfectly with the impact continuum, it is placed towards the top of *Disruption* due to its severity. PS claims independence, and no state involvement has been confirmed, but their position on CIARS suggests Israel would bear responsibility if found to be influencing, supporting, or controlling them.

Chaos Computer Club (CCC) are politically independent but critical of state and corporate practices, shifting their alignment slightly left. As the pioneers of hacktivism, their actions are perhaps the least escalatory, meaning their activities range from *Information* operations to *Nuisance* disruptions, more conventional activities of digital protest. Being that CCC are not aligned with any state, it is unlikely one should be considered responsible for their actions.

CyberAv3ngers (CA) are anti-Israel with purported links to the IRGC, shifting their alignment right. Their activities include information operations and defacement of Unitronics PLCs in western CNI. Although crude, the attack disrupted water access for 180 people in Ireland for two days, placing it in the lower end of *Destabilization*. Their host state is unclear, but CIARS indicates responsibility would fall on a state found to be influencing, supporting, or controlling them.

Finally, NoName057(16) is a pro-Russian group, shifting their alignment right. Their activities, from disinformation campaigns to DDoS attacks, fall between *Nuisance* and *Information*. While claiming independence, their openly communicated alignment with Russian geopolitical interests suggests a deeper connection. Although their actions are not extreme, their alignment places them within the second inner-most ring, meaning their host state would bear responsibility if found to be supporting or controlling them.

As the mappings illustrate, CIARS visually represents and conceptually clarifies the complexity of hacktivism, highlighting its nuances and patterns of activity. It places the focus of such attacks and threat actors back onto the impact of their actions and their politico-ideological alignment, rather than singular terms. This is particularly important if hacktivism continues to escalate and intensify as singular terms will struggle to keep up, whereas CIARS will remain dynamic.

### The implications of escalatory hacktivism

Section 3 confirms the significant shifts in hacktivism, marked by escalatory trends including increased state alignment and evolving impacts. Traditionally situated within the lower-impact range of *Information* and *Nuisance* on the spectrum, contemporary hacktivist activities are increasingly transcending those boundaries. These activities now extend to *Disruption* and even *Destabilization*, with tendencies toward cyber-physical impacts targeting critical national infrastructure (CNI). Kostyuk, Powell, and Skach (2018) highlight how cyber activities can escalate conflicts as they ascend the 'cyber escalation ladder', driven by the interplay of state and non-state actors in increasingly complex geopolitical environments. This evolution carries profound implications for the international security landscape and cyber conflict norms.

First, the increasing alignment of hacktivist agendas with state narratives and objectives creates ambiguity in attribution and response. Attribution and the implications for state responsibility in cyber conflicts remain significant challenges (Banks, 2021). States leveraging hacktivists as proxies or tolerating their activities further complicate identifying the true origin of attacks, undermining traditional deterrence models and complicating diplomatic responses. Reinhold and Reuter (2023) highlight how misattribution due to limited digital evidence exacerbates this issue, potentially leading to escalations rooted in incorrect assumptions of intent or culpability.

Second, the targeting of OT within CNI marks a potentially dangerous escalation, demonstrating hacktivists' willingness to cause cyber-physical impact (Schneider, Schechter, & Shaffer, 2023). While such attacks may fall short of causing harm to life presently, their intent signals a trajectory toward more impactful disruptions. Moreover, the gap between hacktivist intent and achieved results underscores the growing ambition to influence geopolitical outcomes. While current impacts may be limited to temporary disruptions, the intent to destabilize CNI reflects a trajectory toward more severe consequences. Acton (2020) underscores the broader implications of such escalations, noting how unintended consequences, such as cyber operations spilling into nuclear or strategic domains, pose significant risks to global stability. As capabilities improve, potentially through state support, the convergence of intent and impact becomes a more pressing concern, demanding proactive measures to mitigate future risks. This escalatory pattern, therefore, blurs distinctions between hacktivism, cybercrime, and state-sponsored cyber warfare, necessitating a reconsideration of how states address these hybrid threats.

Third, host states face unique challenges. In cases where hacktivists operate independently or without explicit state sanction, host states risk reputational damage, diplomatic fallout, or even escalation if their territory is implicated in attacks on another nation. While some states may benefit from the plausible deniability provided by non-state actors, this strategy risks international backlash. If actors are linked to the state, directly or indirectly, the state might be held accountable for violations of international law, including targeting civilian infrastructure or committing disproportionate attacks. Such activities raise critical questions about state responsibility and the adequacy of existing frameworks to mitigate unauthorized actions by non-state actors (Dinniss, 2013). Clearer attribution mechanisms and policies are required to distinguish state-endorsed activities from rogue operations, reducing the likelihood of inadvertent escalations. Lonergan and Schneider (2023) highlight how evolving state strategies, such as persistent engagement and forward defence, reshape the cyber conflict landscape. While these strategies aim to disrupt adversaries proactively, they may inadvertently create a more contested environment, where the actions of non-state actors intersect with state objectives, potentially escalating tensions in the absence of clear international norms.

Finally, the increasing alignment of hacktivists with state objectives, whether through direct control, tacit support, or ideological alignment, exposes the limits of traditional categorizations. As Kostyuk, Powell, and Skach (2018) highlight, understanding the progression of cyber activities and their potential to provoke strategic escalations requires nuanced frameworks that account for hybrid threats. The observed activity necessitates a more dynamic approach, such as CIARS proposed in this paper, to better assess the risks posed by these actors.

**Addressing escalatory hacktivism**

It is clear from this work that the escalation and intensification of hacktivist tactics are a growing threat. As groups increasingly target OT, particularly in CNI, the chance of cyber-physical and widespread civilian impact increases with it. Below, we offer considerations for the victims, observers, and even hosts of escalatory hacktivist groups.

Strengthening defences and resilience against the trajectory of escalatory hacktivism is essential for all states. This, of course, means a focus on cyber-physical systems, in particular CNI such as water and power. NIS2 (European Union, 2022) does much to achieve this for EU member states and other participating countries. However, many of the vulnerabilities present in such facilities stem from choices made in the supply chain; predominantly from device manufacturers, but also system integrators, maintainers, and other third-parties (Green, et al., 2021), who are seldom incentivized to improve their practices. This asymmetry of regulatory responsibility puts pressure on CNI organizations, who in defending their facilities, must compensate for vulnerabilities introduced by their suppliers. More focus should be put on holding such organizations to account, targeting the root cause of vulnerabilities before they get integrated into CNI facilities.

Clear international frameworks are needed to address the accountability of states linked to escalatory hacktivism. Mechanisms such as public disclosure of evidence linking state-aligned hacktivist activities to specific states can hold these states accountable and apply diplomatic pressure. States should also adopt proportionate response protocols, targeting state-linked hacktivist operations while avoiding escalation, and pursue diplomatic efforts to establish norms prohibiting escalatory hacktivism and building coalitions to counter it. However, diplomacy remains challenging, as the independence of hacktivist groups and the difficulty of attribution complicate efforts to hold states accountable. Multilateral partnerships, including intelligence-sharing frameworks, can enhance collective resilience against hacktivist threats. Our research suggests that public and private sector organizations need to be more cognizant of the specific risks stemming from the increased prevalence of hacktivists targeting OT and CNI, and that this is not just being perpetrated by state actors anymore. Furthermore, public-private collaboration is also vital, as it fosters better intelligence sharing, coordinated defences, and enhanced capacity-building, particularly in low-resource states.

As briefly discussed by Maurer (2018), host states face distinct challenges due to the potential for alignment between hacktivist groups and their national interests. These states must carefully manage their relationships with hacktivist groups to avoid unintended escalations or diplomatic fallout. Host states should publicly distance themselves from independent hacktivist actions while adhering to international norms and publicly discouraging their actions, mitigating risks of perceived complicity. When alignment exists, host states must communicate clearly to ensure hacktivist actions align with national interests without undermining stability.

Host states' direct support for hacktivist groups, such as funding or operational guidance, should be avoided, as it risks an international response or even further escalation. Courting escalatory hacktivists can lead to host states potentially losing control of them and taking responsibility for impacts that they were not prepared for. Instead, host states should formalize state-controlled cyber capabilities to achieve national objectives without relying on potentially uncontrollable escalatory civilian hacktivists.

CIARS helps policymakers by providing a structured approach to analyse hacktivist activities based on their impact, alignment, and state responsibility. It enables the differentiation between low and high-impact cyber-attacks, clarifies whether actors are state-aligned or independent, and identifies levels of state responsibility. This supports policymakers in prioritizing responses, ensuring proportionality, and developing strategies that balance defence, accountability, and international collaboration.

By taking these considerations into account, states can collectively begin to address the immediate and long-term challenges posed by escalatory hacktivism, fostering greater stability and cooperation in cyberspace.

## Conclusion

This paper has demonstrated the challenges of escalatory hacktivism in contemporary cyber conflict, transitioning from protest-driven actions to increasingly impactful attacks on operational technology (OT) and critical national infrastructure (CNI). As these activities align more closely with state narratives, hacktivism has shifted from its independent origins to a domain of cyber operations that often blurs the lines between activism, cybercrime, and state-linked activities such as cyber warfare.

The introduction of the Cyber Impact-Alignment-Responsibility Spectrum (CIARS) offers a way to assess the nuances of modern hacktivism. By categorizing actors based on their politico-ideological alignment and impact severity, and suggesting thresholds of host state responsibility, CIARS provides a dynamic tool to evaluate and address the challenges posed by this evolving threat. The spectrum not only helps policymakers contextualize hacktivist actions but also clarifies the role of states in enabling or mitigating such activities, whether through direct control, tacit support, or failure to act.

The escalation in hacktivist tactics—from low-level disruptions to attacks targeting OT within CNI—represents a concerning trend. These attacks reflect an intent to achieve geopolitical objectives through destabilization and disruption. The implications for victim states may be severe, ranging from disruption of public services to broader destabilization and harm to life. Host states, meanwhile, face growing scrutiny over their role in tolerating, influencing, or supporting hacktivist activities. These states risk diplomatic fallout and reputational damage if implicated in actions that contravene international norms or expectations. For observer states, the challenge lies in fostering stability and mitigating the risks posed by hacktivist-aligned geopolitical shifts.

Policy responses must evolve to address these challenges effectively. Strengthening defence and resilience in cyber-physical systems, particularly CNI, is critical for all states. Collaborative efforts, such as public-private partnerships and multilateral intelligence-sharing agreements, are essential for enhancing collective defences. Attribution mechanisms must be refined to avoid misattribution and unnecessary escalation, while proportionate responses should be carefully calibrated to minimize conflict intensification.

For host states, formalizing state-controlled cyber capabilities provides a structured alternative to potentially uncontrollable and escalatory hacktivist operations that may provoke further, unintended escalations. Clear communication and adherence to international norms are vital to distancing themselves from independent hacktivist actions.

As hacktivism continues to evolve, the need for dynamic and context-sensitive approaches to classification and response becomes increasingly evident. CIARS provides a foundation for understanding the interplay of impact, alignment, and state responsibility in hacktivist actions, offering a basis for tailored and effective policy interventions. By integrating these considerations into national and international strategies, states can better navigate the challenges posed by escalatory hacktivism, ensuring stability and security in an era of intensifying cyber conflict.

**References**

2600; Chaos Computer Club; Cult of the Dead Cow; !HISPAHACK; L0pht Heavy Industries; Phrack; Pulhas. (1997, 01 7). *LoU STRIKE OUT WITH INTERNATIONAL COALITION OF HACKERS*. Retrieved 12 23, 2024, from https://cultdeadcow.com/news/statement19990107.html

Acton, J. M. (2020). Cyber Warfare & Inadvertent Escalation. *Daedalus*, 133-149. doi:https://doi.org/10.1162/daed_a_01794

Banks, W. (2021). Cyber Attribution and State Responsibility. *International Law Studies, 97*, 1039-1072. Retrieved 01 03, 2025, from https://digital-commons.usnwc.edu/ils/vol97/iss1/43/

Bertram, S. K. (2017). 'Close enough'–The link between the Syrian Electronic Army and the Bashar al-Assad regime, and implications for the future development of nation-state cyber counter-insurgency strategies. *Contemporary Voices: St Andrews Journal of International Relations*, 2-17. doi:https://doi.org/10.15664/jtr.1294

Betlej, A. (2023). The Faces of Hacktivism by the Anonymous Collective in the Context of Russian War Against Ukraine. Comparison Between 2014 and 2022. *Journal of Security and Sustainability Issues*, 143-152. doi:https://doi.org/10.47459/jssi.2023.13.14

Caldwell, T. (2015). Hacktivism goes hardcore. *Network Security*, 12-17. doi:https://doi.org/10.1016/S1353-4858(15)30039-8

Chaos Computer Club. (n.d.). *Hacker Ethics*. Retrieved 12 23, 2024, from https://www.ccc.de/en/hackerethics

Chng, S., Lu, H. y., Kumar, A., & Yau, D. (2022). Hacker types, motivations and strategies: A comprehensive framework. *Computers in Human Behavior Reports*, 100167.

CISA. (2024). *IRGC-Affiliated Cyber Actors Exploit PLCs in Multiple Sectors, Including US Water and Wastewater Systems Facilities.* Cybersecurity advisory. Retrieved 01 03, 2025, from https://www.cisa.gov/news-events/cybersecurity-advisories/aa23-335a

Denning, D. E. (2001). Activism, Hacktivism, and Cyberterrorism: The Internet as a Tool for Influencing Foreign Policy. In J. Arquilla, *Networks and netwars: The future of terror, crime, and militancy.*

Derbyshire, R. (2022). *Anticipating Adversary Cost: Bridging the Threat-Vulnerability Gap in Cyber Risk Assessment.* Lancaster University.

Derbyshire, R., Green, B., Van der Walt, C., & Hutchison, D. (2024). Dead Man's PLC: Towards Viable Cyber Extortion for Operational Technology. *Digital Threats: Research and Practice, 5*(3), 1-24. doi:https://doi.org/10.1145/3670695

Dinniss, H. H. (2013). Participants in Conflict – Cyber Warriors, Patriotic Hackers and the Laws of War. In D. Saxon, *International Humanitarian Law and the Changing Technology of War* (pp. 251-278). Leiden: Martinus Nijhoff Publishers. doi:https://doi.org/10.1163/9789004229495_013

European Union. (2022, 12 27). Directive (EU) 2022/2555 of the European Parliament and of the Council of 14 December 2022 on Measures for a High Common Level of Cybersecurity Across the Union. *Official Journal of the European Union L 333*, 80-131. Retrieved from https://eur-lex.europa.eu/eli/dir/2022/2555/oj

Goode, L. (2015). Anonymous and the Political Ethos of Hacktivism. *Popular Communication*, 74-86. doi:https://doi.org/10.1080/15405702.2014.978000

Green, B., Derbyshire, R., Krotofil, M., Knowles, W., Prince, D., & Suri, N. (2021). PCaaD: Towards automated determination and exploitation of industrial systems. *Computers and Security, 110*. doi:https://doi.org/10.1016/j.cose.2021.102424

Healey, J. (2011). The spectrum of national responsibility for cyberattacks. *The Brown Journal of World Affairs*, 57-70.

IBM. (2019). *X-Force Threat Intelligence Index 2019.* IBM.

International Committee of the Red Cross. (2024). *Eight rules for "civilian hackers" during war, and four obligations for states to restrain them.* Retrieved from https://www.icrc.org/en/article/8-rules-civilian-hackers-during-war-and-4-obligations-states-restrain-them

Kalinowski, M. (2022, 08 17). *Predatory Sparrow operation against Iranian steel maker (2022)*. Retrieved 01 03, 2025, from https://cyberlaw.ccdcoe.org/wiki/Predatory_Sparrow_operation_against_Iranian_steel_maker_(2022)

Karagiannopoulos, V. (2021). A Short History of Hacktivism: Its Past and Present and What Can We Learn from It. In T. Owen, & J. Marshall, *Rethinking Cybercrime: Critical Debates* (pp. 63-86). doi:https://doi.org/10.1007/978-3-030-55841-3_4

Kostyuk, N., Powell, S., & Skach, M. (2018). Determinants of the Cyber Escalation Ladder. *The Cyber Defense Review*, 123-134.

Lonergan, E. D., & Schneider, J. (2023). The Power of Beliefs in US Cyber Strategy: The Evolving Role of Deterrence, Norms, and Escalation . *Journal of Cybersecurity*. doi:https://doi.org/10.1093/cybsec/tyad006

Maurer, T. (2015). Cyber Proxies and the Crisis in Ukraine. In K. Geers, *Cyber War in Perspective: Russian Aggression Against Ukraine* (pp. 79-86). Estonia: NATO CCD COE Publications.

Maurer, T. (2018). *Cyber Mercenaries.* Cambridge University Press. doi:https://doi.org/10.1017/9781316422724

McFadden, P. (2024). Chancellor of the Duchy of Lancaster's speech. *NATO Cyber Defence Conference at Lancaster House.* London: Cabinet Office. Retrieved from https://www.gov.uk/government/speeches/chancellor-of-the-duchy-of-lancasters-speech-to-the-nato-cyber-defence-conference

Menn, J. (2019). *Cult of the Dead Cow: How the Original Hacking Supergroup Might Just Save the World.* Pulic Affairs.

OneFist, V. o. (2024, 09 01). *Twitter post*. Retrieved 01 03, 2025, from https://x.com/SpoogemanGhost/status/1830029859726500124

Reinhold, T., & Reuter, C. (2023). Preventing the escalation of cyber conflicts: towards an approach to plausibly assure the non-involvement in a cyberattack. *Zeitschrift für Friedens-und Konfliktforschung*, 31-58. doi:https://doi.org/10.1007/s42597-023-00099-7

Romagna, M. (2024). Social opportunity structures in hacktivism: Exploring online and offline social ties and the role of offender convergence settings in hacktivist networks. *Victims & Offenders*.

Romagna, M., & Leukfeldt, E. R. (2024). Hacktivism: From loners to formal organizations? Assessing the social organization of hacktivist networks. *Deviant Behavior*, 1-23.

Samuel, A. W. (2004). *Hacktivism and the future of political participation.* Harvard University.

Schneider, J., Schechter, B., & Shaffer, R. (2023). Hacking Nuclear Stability: Wargaming Technology, Uncertainty, and Escalation. *International Organization*, 633-667.

Selck-Paulsson, D., & Gibney, B. (2024, December 5). Exploring the Intersection of Cyber Activism and State-sponsored Operations. *Security Navigator 2025*, pp. 86-95.

Smith, M. W., & Dean, T. (2023). The Irregulars: Third-Party Cyber Actors and Digital Resistance Movements in the Ukraine Conflict. *15th International Conference on Cyber Conflict: Meeting Reality (CyCon)* (pp. 103-119). Tallinn: IEEE. doi:https://doi.org/10.23919/CyCon58705.2023.10182061

Svantesson, D. J. (2023). Regulating a "Cyber Militia" - Some Lessons from Ukraine, and Thoughts about the Future. *Scandinavian Journal of Military Studies*, 86-101. doi:https://doi.org/10.31374/sjms.195

Zafra, D. K., Lunden, K., & Brubaker, N. (2023). *We (Did!) Start the Fire: Hacktivists Increasingly Claim Targeting of OT Systems.* Mandiant. Retrieved 01 03, 2025, from https://cloud.google.com/blog/topics/threat-intelligence/hacktivists-targeting-ot-systems

Zafra, D. K., Wahlstrom, A., Sadowski, J., Palatucci, J., Vaumann, D., & Nazario, J. (2024, June 27). *Global Revival of Hacktivism Requires Increased Vigilance from Defenders.* Mandiant. Retrieved 01 03, 2025, from https://cloud.google.com/blog/topics/threat-intelligence/global-revival-of-hacktivism

Zettl-Schabath, K. (2023). *Staatliche Cyberkonflikte: Proxy-Strategien von Autokratien und Demokratien im Vergleich.*